\documentclass[12pt]{article}
\usepackage[dvips]{graphicx}
\usepackage{latexsym}
\def\Comment#1{}

\newcommand{\bean}{\begin{eqnarray*}}
\newcommand{\eean}{\end{eqnarray*}}

\newcommand{\gapproxeq}{\lower
.7ex\hbox{$\;\stackrel{\textstyle >}{\sim}\;$}}
\newcommand{\lapproxeq}{\lower
.7ex\hbox{$\;\stackrel{\textstyle <}{\sim}\;$}}

\newcommand\lsim{\mathrel{\rlap{\lower4pt\hbox{\hskip1pt$\sim$}}
    \raise1pt\hbox{$<$}}}
\newcommand\gsim{\mathrel{\rlap{\lower4pt\hbox{\hskip1pt$\sim$}}
    \raise1pt\hbox{$>$}}}
\newcommand{\ba}{\begin{array}}
\newcommand{\ea}{\end{array}}

\newcommand{\be}{\begin{equation}}
\newcommand{\ee}{\end{equation}}
\newcommand{\bear}{\begin{eqnarray}}
\newcommand{\eear}{\end{eqnarray}}
\newcommand{\tab}{\hspace*{0.5cm}}

\newcommand{\rvac}{\,|0\rangle}

\newcommand{\ket}{\,\rangle}
\newcommand{\bra}{\langle \,}
\newcommand{\eqn}[1]{(\ref{#1})}
\newcommand{\cO}{{\cal O}}
\newcommand{\bel}[1]{\be\label{#1}}
\newcommand{\mL}{\mathcal{L}}

\newcommand{\mM}{\mathcal{M}}

\newcommand{\mF}{\mathcal{F}}
\newcommand{\mT}{\mathcal{T}}

\newcommand{\mG}{\mathcal{G}}
\newcommand{\Frac}[2]{\frac{\displaystyle #1}{\displaystyle #2}}
\newcommand{\Int}{\displaystyle{\int}}
\begin{document}
\thispagestyle{empty}
\begin{titlepage}
\begin{center}
\hfill IFIC/02$-$14 \\ 
\hfill FTUV/02$-$0822 \\
\vspace*{2.75cm} 
\begin{Large}
{\bf Rho Meson Properties in the Chiral Theory Framework}
\\[2.4cm]
\end{Large}
{\sc J.J. Sanz-Cillero } and \ {\sc A. Pich }\\[0.8cm]

{\it Departament de F\'\i sica Te\`orica, IFIC, Universitat de Val\`encia -
CSIC\\
 Apt. Correus 22085, E-46071 Val\`encia, Spain }\\[0.5cm]
\vspace*{2cm}
\begin{abstract}
\noindent
We  study the mass, width and couplings  of the lightest 
resonance multiplet with  $I\, (J^{PC})=1\, (1^{--})$ quantum numbers.
Effective field theories based on chiral symmetry are employed in order to
describe the form factor associated with the two-pseudoscalar
matrix element of the QCD vector current. 
The bare poles of the intermediate resonances are regularized
through a Dyson-Schwinger-like summation. 
We  explore the role of the resonance width in physical observables 
and make a coupled-channel analysis of final-state interactions. This provides
many interesting properties, as the pole mass 
$M_\rho^{\mbox{\tiny pole}}= 764.1\pm 2.7\, {}^{+4.0}_{-2.5}$ MeV.
At energies  $E\gsim 1$~GeV,  a second  $1\, (1^{--})$ resonance 
multiplet is considered in order  to  describe the data in a more consistent
way. From the phenomenologically extracted resonance couplings,
we obtain the chiral coupling 
$L_9^r(\mu_0)= (7.04\pm 0.05\, {}^{+0.19}_{-0.27})\cdot 10^{-3}$, 
at $\mu_0=770$ MeV, and show how the  running with the scale $\mu$ 
affects the resonance parameters.
A $1/N_C$ counting is adopted in this work and the consistency of the
large--$N_C$ expansion is tested.
Finally, we make an estimation of the contribution from diagrams with resonances
in crossed channels.   
\end{abstract}
\end{center}
\vfill
\eject
\end{titlepage}

\pagenumbering{arabic}

\parskip12pt plus 1pt minus 1pt
\topsep0pt plus 1pt
\setcounter{totalnumber}{12}

\section{Introduction}
\tab
It has become evident that Quantum Chromodynamics (QCD) 
is the correct theory which describes hadronic processes~\cite{QCDPich}.
In the high-energy region ($E\gg 1$ GeV)  
the theory accepts a perturbative description and, accordingly,   
many calculations up to several orders in the perturbative
expansion parameter $\alpha_s$ 
have been performed. These theoretical results 
have been successfully tested in many high-energy experiments. 
Nevertheless,  
since the running coupling constant $\alpha_s(\mu)$ increases as
the energy decreases, the perturbative expansion in powers of
$\alpha_s$ breaks down at energies $E\sim 1$ GeV. In this paper 
the problem of describing the $E\lsim 1$ GeV region by   
employing effective theories of QCD~\cite{EFT,libro Dobado} will be analyzed.

When we study processes at energies much lower than the  heavy quark masses,
the degrees of freedom corresponding to heavy quarks decouple~\cite{desac}
and QCD, with only the light quark fields, yields a proper description.
In the massless limit, the  QCD lagrangian shows chiral symmetry:  
the left-handed and right-handed quark fields can be rotated independently
under the $SU(n_f)_L\otimes SU(n_f)_R$ flavour 
chiral group, where $n_f$ is the number of light
quarks. The symmetry is spontaneously broken to the 
$SU(n_f)_V$ subgroup and $n_f^2-1$ massless 
Nambu-Goldstone bosons appear, associated with the broken generators.
Nonetheless, as the light quark QCD
lagrangian has small non-zero mass terms, chiral symmetry is also 
broken explicitly  and the Nambu-Goldstone bosons gain small masses.  
These bosons have $J^P=0^-$ and are 
identified with the triplet of pions, in the $SU(2)$ case,
and the ($\pi, \ K, \ \eta_8$) octet of light pseudoscalars for $SU(3)$.

The low-energy chiral effective field theory
describing the dynamics of the lightest pseudoscalar multiplet 
was first developed for the $SU(2)_L\otimes SU(2)_R$ 
symmetry group~\cite{chpt1loop},   
and was later generalized to the three flavour 
$SU(3)_L\otimes SU(3)_R$ case~\cite{chptms}.
We will use the latter in order to include kaon interactions
in our study.
Chiral Perturbation Theory
($\chi$PT)~\cite{chpt1loop,chptms,WE:79,EC:95,PI:95} describes  
the physical low-energy amplitudes as an expansion in powers 
of quark masses and momenta over a characteristic chiral scale 
$\Lambda_\chi \simeq 4\pi f_\pi \sim 1$ 
GeV, with $f_\pi= 92.4$~MeV the pion decay constant.  

The expansion in powers of  momenta over $\Lambda_\chi$ deteriorates as   
the energy of the process is increased and,  
in order to reach the relevant   accuracy, one needs to add higher and higher 
chiral orders to the $\chi$PT lagrangian.  In the resonance region one must
introduce a different effective field theory with explicit massive fields to
describe the degrees of freedom associated with the mesonic resonances.  
In the eighties, Gasser and Leutwyler worked out 
an $SU(2)_L\otimes SU(2)_R$ lagrangian   
describing the pions and the vector resonance
$\rho(770)$~\cite{chpt1loop}. Later on this  work was extended to the $n_f=3$ 
case~\cite{the role}, developing 
the Resonance Chiral Theory (R$\chi$T). Further studies
on the R$\chi$T and $\chi$PT lagrangians constrained the resonance chiral couplings, 
employing the QCD short-distance behaviour
of appropriate Green functions~\cite{spin1fields}. 

Once the resonance fields are explicitly included in the
effective lagrangian, the chiral counting becomes ineffective because the
masses of these resonances are of the same order than the
chiral characteristic scale $\Lambda_\chi$. However, 
an expansion of QCD and its low-energy effective field theory
in powers of $1/N_C$, with $N_C$ the number of quark colours,
appears to be suitable \cite{NC}. In the large--$N_C$ limit, 
the hadronic description reduces to  
tree-level processes without hadron loops. As 
the $1/N_C$ expansion seems to yield a proper description of 
$N_C=3$ \ QCD, it seems also appropriate to expand 
the R$\chi$T results in powers of $1/N_C$ \cite{PI:02}.
To a certain extent, this reduces to just counting the number of loops.     

At leading order (LO) in $1/N_C$, R$\chi$T yields a good description of many
phenomena. However it fails when the energy approaches the bare mass 
of a resonance. This situation is common to every unstable
propagating state in a Quantum Field Theory when its 
propagator turns on-shell. It is solved by the Dyson-Schwinger summation of
one particle insertion blocks (1PI), which provides the unstable particles 
with an imaginary absorptive part in the resonance propagator. 
This summation must also be done in R$\chi$T, with some  
prescriptions, but essentially in the same way. In 
Refs.~\cite{GyP,anchura Jorge} the $\rho$-channel was
studied and an appropriate off-shell width for the $\rho(770)$ 
resonance was obtained.

In this paper we continue the work put forward in Ref.~\cite{anchura Jorge},  
extending it to a coupled-channel analysis. 
We will study the vector form factor
(VFF)~\cite{GyP,anchura Jorge,PP:01,Palomar,Yndurain} 
and overview the correlator of two
QCD vector currents and the corresponding partial-wave scattering amplitude. 
It will be shown that our coupled-channel description of the resonance 
width agrees with the one in Ref.~\cite{anchura Jorge},   
obtained with a single-channel treatment. From 
the Dyson-Schwinger summation, we find that
the rescattering dresses the bare propagator in a universal
way. The induced correction only depends 
on the intermediate 1PI blocks and not on the final or initial
states of the process.  

We briefly describe the basic ingredients of the R$\chi$T effective action
in Section~2.
The Dyson-Schwinger analysis of the different observables
is performed in Section 3.
In Section 4 the obtained results are matched to the $\cO(p^4)$ 
$\chi$PT description and compared with the data in Section 5. The $\chi$PT coupling
$L_9^r(\mu)$ is calculated here and a test of the $1/N_C$ expansion is also
performed. The small corrections induced by resonance exchanges in the
$t$--channel are estimated in Section 6.
Our conclusions are finally given in Section 7. 
Some technical details have been relegated to the Appendices.
In particular, a generalized formal summation of diagrams with two-body
topologies is presented in Appendix~C.   

\section{Resonance Chiral Theory}
\tab 
We will work with the $SU(3)$ octet of light pseudoscalar bosons,  
interacting through the $\cO(p^2)$ $\chi$PT lagrangian~\cite{chptms,the role}:
\bel{eq.L2}
\mL_{\mbox{\tiny $\chi$PT}}^{(2)}\;=\;\Frac{f^2}{4}\;
\bra u^\mu u_\mu\, +\, \chi_+ \ket\, ,  
\ee
where $\bra ... \ket$ is short for the trace over flavour matrices
and $f\approx f_\pi$ is the pion decay constant at lowest order.
The tensors 
$u^\mu = i [ u^\dagger (\partial^\mu -ir^\mu)u -u(\partial^\mu -il^\mu)
u^\dagger]$ and 
$\chi_+ = u^\dagger \chi u^\dagger + u \chi^\dagger u$ 
are functions of the left, right and scalar external
sources $l^\mu,\ r^\mu$ and $\chi$~\cite{chpt1loop,chptms}.   
The pseudoscalar fields
\be
 \Phi  =
 \left( \begin{array}{ccc}
\frac{\pi^{0}}{\sqrt{2}} + \frac{\eta_8}{\sqrt{6}} & \pi^{+} & K^{+} \\
\pi^{-} & -\frac{\pi^{0}}{\sqrt{2}} + \frac{\eta_8}{\sqrt{6}} & K^{0} \\
K^{-} & \overline{K}^{0} & - \frac{ 2 \eta_8}{\sqrt{6}} 
\end{array} \right)
\ee
are parameterized through the $SU(3)$ matrix \
$u\equiv \exp{\left[\frac{i}{f\sqrt{2}}\,\Phi \right]}$. 

The interactions of the Nambu-Goldstone bosons with the
lightest multiplet of vector resonances 
\bel{eq.octeterho}
V_{\mu\nu}\, =\, \Frac{\lambda^a}{\sqrt{2}} V_{\mu\nu}^a\, =\,
\pmatrix{                                                       
{1\over\sqrt{2}}\rho^0 + {1\over \sqrt{6}}\omega_{8} 
+ {1\over \sqrt{3}}\omega_{1}
& \rho^+  & K^{*+} 
\cr                                                                 
\rho^-  & - {1\over\sqrt{2}}\rho^0  + 
{1\over \sqrt{6}}\omega_{8 } + {1\over \sqrt{3}}\omega_{1}
& K^{\, *0} 
\cr
K^{*-}  & \bar{K}^{*0}  
& -{2\over \sqrt{6}}\omega_{8 } + {1\over \sqrt{3}}\omega_{1}                        
}_{\mu\nu}  
\ee
are given by~\cite{the role}
\bel{eq.LV}
\mL_{\mbox{\tiny R$\chi$T}}^{V}\, =\,
\Frac{F_V}{2\sqrt{2}}\;\bra\ V_{\mu\nu} f_+^{\mu\nu} \ket
 \, +\, i\; 
\Frac{G_V}{\sqrt{2}}\;  \bra V_{\mu\nu} u^\mu u^\nu \ket\,   , 
\ee 
where  $f_+^{\mu\nu}= u\, F_L^{\mu\nu} u^\dagger + u^\dagger F_R^{\mu\nu} u$, 
with $F_{L,R}^{\mu\nu}$ the field strength tensors of 
the left and right external fields~\cite{chpt1loop,chptms}.  
We use the R$\chi$T lagrangian 
in the antisymmetric formalism provided in~Ref.~\cite{the role}.
It was demonstrated in Ref.~\cite{spin1fields} that this 
antisymmetric description of the vector fields 
is equivalent to the more usual Proca formalism plus the $\cO(p^4)$
$\chi$PT lagrangian with its couplings $L_i$ constrained by the short-distance
QCD behaviour.

In general, one should consider a set of vector resonance multiplets
$V^{(i)}_{\mu\nu}$ with couplings $F_{V_i}$ and $G_{V_i}$. At low energies
($\sqrt{s}\lsim 1.2$ GeV), the lightest multiplet yields 
the dominant contributions. However the tail of the second nonet
may generate sizable corrections which
must also  be  taken into account in the $\sqrt{s}\sim 1$~GeV region.

\section{The Vector Form Factor}
\tab
Let us consider the hadronic matrix element corresponding to the
production of two pseudoscalars with $I=J=1$ through the
charged $\bar{d}\,\gamma^\mu u$ vector current:  
\bel{eq.VFFdef}
\bra P^-(p_1) \, P^0(p_2)\, |\,
\bar{d}\,\gamma^\mu u\,\rvac \; = \;\sqrt{2}  \ \left(
p_1-p_2\right)^\mu \   \mF^{(P)}_{\phantom{0}}(q^2) \, ,
\ee
with $q=p_1+p_2$.  
The label $P$ denotes the pair of pseudoscalars which are 
produced in the final state, either $\pi^-\pi^0$ or $K^-K^0$.
The Lorentz structure is fixed by current conservation in the isospin
limit.

\begin{figure}[t]
\begin{center}
\includegraphics[width=10cm,clip]{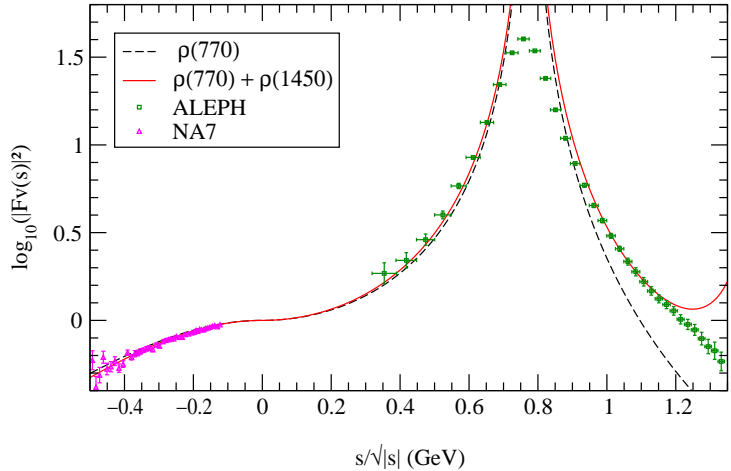}
\caption{\small 
VFF at leading order in $1/N_C$ with one and two 
vector resonances. For the two-resonance case we have adopted the
input parameters $M_{V_1}=775$~MeV, $M_{V_2}=1450$~MeV,
$F_{V_1} G_{V_1}/f^2=1.1$ and $F_{V_2} G_{V_2}/f^2=-0.1$. 
Data from ALEPH~\cite{ALEPH} and NA7~\cite{Amendolia}.}
\label{fig.LO}
\end{center}
\end{figure}

At leading order in $1/N_C$, the vector form factor 
$\mF^{(P)}_0(q^2)$
is easily computed through the diagrams shown in Fig.~\ref{fig.verteff}(a).
We  put together the two $\mF_0^{(P)}(q^2)$ functions 
in the vector 
\bel{eq.inivec}
\vec{\mF}_0(q^2) \;\equiv\;
\left( \ba{c} \mF_0^{(\pi)}(q^2) \\ \mF_0^{(K)}(q^2) \ea \right) 
\; = \; 
\left\{ 1\, +\, \sum_i\;\frac{F_{V_i} G_{V_i}}{f^2} 
\; \Frac{q^2}{M_{V_i}^2-q^2}\right\} \  
\left( \ba{c} 1 \\ -\frac{1}{\sqrt{2}} \ea \right) \, . 
\ee
The requirement that the vector form factor should vanish at
infinite momentum transfer constrains the
resonance couplings at LO in $1/N_C$  
to satisfy the short-distance QCD relation~\cite{spin1fields,PI:02}
\bel{eq:eps}
1\ - \  \sum_i\,\frac{F_{V_i} G_{V_i}}{f^2} \; =\; 0 \, .
\ee
If only one vector multiplet is considered,
then $F_{V_1} G_{V_1}/f^2 = 1$
and one gets the familiar vector-meson dominance expression 
\be 
\vec{\mF}_0(q^2)  \; = \; 
\Frac{M_{V_1}^2}{M_{V_1}^2-q^2}  \  
\left( \ba{c} 1 \\ -\frac{1}{\sqrt{2}} \ea \right) \, . 
\ee
This yields a rather good description of the data
in the region $\sqrt{q^2}\lsim 0.7$~GeV, below the $\rho(770)$ peak.
Chiral loop corrections are subleading in the $1/N_C$ counting and
turn out to be rather small in this case. 
Other resonances can also be included. 
The relevance of the large--$N_C$ expansion 
to approximate the physical vector form factor is clearly seen in
Fig.~\ref{fig.LO},  
either with just one resonance or including a second multiplet. 

The vector couplings are as well constrained in the large--$N_C$ limit 
by the relation
\bel{eq.delta}
\sum_i\, \Frac{2\, F_{V_i} G_{V_i} - F_{V_i}^2}{M_{V_i}^2}\; =\; 0 \, ,
\ee
provided by the short-distance QCD conditions over the 
axial form factor~\cite{spin1fields,PI:02}. 

For the simplest situation with a
single resonance exchange,  the short-distance QCD
constraints yield $F_{V_1} = 2\, G_{V_1} = \sqrt{2}\, f$ 
in the large--$N_C$ limit~\cite{spin1fields}. 
However, since we are going to work at higher orders in $1/N_C$,  
we will leave these couplings free and will test afterwards the deviation of
their experimental values from the large--$N_C$ predictions, 
that we expect to be small.

\subsection{Dyson-Schwinger Summation}
\begin{figure}[t!]
\begin{center}
\includegraphics[height=5cm]{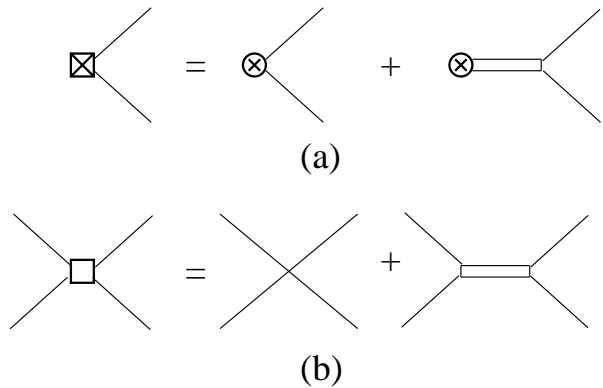}
\caption{\small
Effective vertices for the vector current insertion producing two
pseudoscalars (a) and for the two pseudoscalar scattering (b). 
The first terms come from $\mL_{\mbox{\tiny $\chi$PT}}^{(2)}$
and the second ones from the interaction via
an intermediate resonance due to the $\mL_{\mbox{\tiny R$\chi$T}}^{V}$ 
lagrangian.}
\label{fig.verteff}
\end{center}
\end{figure}

\tab
At energies close to the mass of a resonance we need to know the denominator of
the resonance propagator beyond the  leading, bare, order in $1/N_C$.  What is
usually done is a Dyson-Schwinger summation, as for instance in the QED photon
polarization. That is, summing diagrams composed by
a series of propagator, 1PI block, propagator, \ldots, and so on. This 
summation regularizes the pole of the bare propagator.
It gives a self-energy with its corresponding absorptive part, 
up to the perturbative order employed for the 1PI block.  
In R$\chi$T, however,  
at the same order than the resonance-exchange contribution there is also  
a local interaction from the $\mL_{\mbox{\tiny $\chi$PT}}^{(2)}$ lagrangian. 
The Dyson-Schwinger summation must be then slightly modified.
One constructs effective current vertices and effective scattering vertices
\cite{anchura Jorge}, by adding the contribution from intermediate resonance 
exchanges in the $s$--channel to the local $\chi$PT interaction  
$\mL_{\mbox{\tiny $\chi$PT}}^{(2)}$.
Both contributions are of the same order in the $1/N_C$ counting. 
These effective vertices, shown in Fig.~\ref{fig.verteff}, are
independent of the explicit formulation adopted for the spin--1
fields~\cite{spin1fields,anchura Jorge}. 
If we use the Proca formulation we have to
take into account the local interaction from the $\cO(p^4)$ $\chi$PT lagrangian as
it is described in Ref.~\cite{spin1fields}. The inclusion of the local vertices
is not important  on the resonance peak but it turns to be relevant away from
it. 

\begin{figure}[t!]
\begin{center}
\includegraphics[width=10cm]{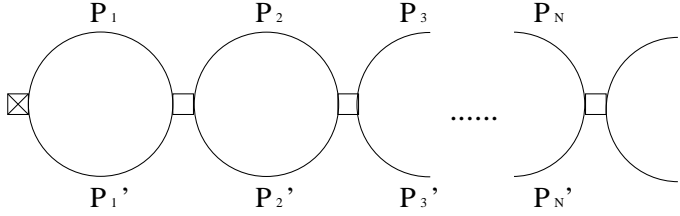}
\caption{\small{Diagrammatic summation at $N$ loops.}}
\label{fig.Dyson}
\end{center}
\end{figure}

For the moment, we are only interested in the imaginary part of the 
self-energy. Therefore, we will concentrate in the sum over diagrams
with absorptive cuts. For the range of energies we
are interested, the most relevant contributions come from
intermediate states with two pseudoscalars;
states with a higher number of particles being
suppressed by phase space and chiral counting.  
Thus, we are going to sum diagrams\footnote{
This diagrammatic construction solves the Bethe-Salpeter equation~\cite{BSE}
in an iterative way. The effective vertices provide the
corresponding ``potentials'' at LO in $1/N_C$.}
constructed with
an initial effective current insertion connected to an effective 
scattering vertex through a two-pseudoscalar loop. 
The pair of outgoing pseudoscalars from the scattering vertex are again
connected to another effective scattering vertex through another 
two-pseudoscalar loop, and so on, as it can be seen in Fig.~\ref{fig.Dyson}. 

The off-shell effective current vertex shows the momentum structure
\bel{eq:offshell} 
\vec{V}_0^{\mu} \; = \; \sqrt{2}\,
\left[ \  \vec{\mF}_0\,  P_T^{\mu\nu} \ 
+ \  \vec{\mF'}_0\,  P_L^{\mu\nu} \ \right] 
\  (p_1-p_2)^{\nu} \, ,
\ee
with\ $P_T^{\mu\nu}=g^{\mu\nu}-\frac{q^\mu q^\nu}{q^2}$\ and\ 
$P_L^{\mu\nu}=\frac{q^\mu q^\nu}{q^2}$\ the usual transverse and longitudinal
Lorentz projectors. 
In the isospin limit, the second term with $\vec{\mF'}_0$ vanishes
when the outgoing pseudoscalars are both on-shell.
Notice that this off-shell function depends on the adopted 
parameterization of the fields, but the final on-shell amplitude
does not depend on it.

When the current insertion $\vec{V}_0^{\mu}$  
is connected to a successive number of loops and 
effective scattering vertices one gets 
$\vec{V}_N^{\mu}=\sqrt{2} \left[ \vec{\mF}_N P_T^{\mu\nu} + 
\vec{\mF'}_N P_L^{\mu\nu} \right] (p_1-p_2)^{\nu}$, 
where $N$  is  the number of intermediate loops in the
diagrammatic chain shown in Fig.~\ref{fig.Dyson}. Thus the momentum structure
remains.   
Inductively,  from $N$ to $N+1$ loops 
we can observe  the linear recurrence 
$\mF_{N+1}^{(i)}=\sum_j \,\mM^{ij}\,\mF_N^{(j)}$, where
$i=1$ stands for $\pi\pi$ and $i=2$ for $K\overline{K}$.
This feature can be expressed in the matrix form
\bel{eq.linealV}
 \vec{\mF}_{N+1}\, =\, \mM\; \vec{\mF}_N\, =\,\mM^2\;\vec{\mF}_{N-1}
 \, =\, \cdots \, =\, \mM^{N+1}\; \vec{\mF}_0 \, , 
\ee
with  $\vec{\mF}_N$ the vector form factor at $N$ loops. 
The $2\times 2$ recurrence matrix takes the form
\bel{eq.MTLO}
\mM  \ \ = \ - \ 
\Sigma^{-1}\, T_{\mbox{\tiny LO}}^s\,\Sigma^{-1} \  (192\,\pi\, B_{22} )  
\,  , 
\label{eq.recuV} 
\ee
with the diagonal matrix  
$\Sigma=\ $diag$\, \left(\sigma_\pi,\ \sigma_K\right)$, being 
$\sigma_{P}=\sqrt{1-4\, m_P^2/q^2}$.  
The matrix
\bel{eq.TLO11}
T_{\mbox{\tiny LO}}^s \; = \;   \Frac{q^2}{96 \pi   f^2} \ 
\left\{\, 1 \, +\, 
\sum_i\;
\Frac{2\, G_{V_i}^2}{f^2}\; 
\Frac{q^2}{M_{V_1}^2-q^2}\,\right\} \;\;\Sigma\;\;
  \left( \ba{cc} 1 & -\frac{1}{\sqrt{2}} 
  \\[10pt]
  -\frac{1}{\sqrt{2}} & \frac{1}{2} \ea \right)\;\;\Sigma  \, ,
\ee 
is the $s$--channel partial-wave scattering amplitude  
with $I=J=1$,  at LO in $1/N_C$ [Fig.~\ref{fig.verteff}(b)].
The  diagrams with resonances in the crossed channels produce a tiny
contribution which will be  taken into account in Section 6.  
We can also observe in Eq.~\eqn{eq.MTLO} 
the diagonal matrix  
$B_{22}=\ $diag$\,\left(B_{22}^{(\pi)},\ B_{22}^{(K)}\right)$,
with the two-propagator Feynman integral $B_{22}^{(P)}$ 
given in Appendix~A.

Summing the result in Eq.~\eqn{eq.linealV} for any number of loops, 
one gets a geometrical series which can be easily handled:
\bel{eq.V0aV}
\vec{\mF} \; = \;\sum_{N=0}^{\infty}\, \vec{\mF}_N \; = \; 
\left( \sum_{N=0}^{\infty}\, \mM^N \right)\; \vec{\mF}_0  
\; = \;  \left(1-\mM\right)^{-1}\, \vec{\mF}_0\; =\;
\Frac{1}{1-\mbox{tr}(\mM)}\; \vec{\mF}_0 \, .  
\ee
The last identity is not trivial.  
The $\Sigma^{-1} T^s_{\mbox{\tiny LO}} \Sigma^{-1}$ matrix is proportional to a
dimension-one projector and $\vec{\mF}_0$ is eigenvector of this projector.
Thus,  
$\mM^N$ acting over $\vec{\mF}_0$ reproduces again the vector $\vec{\mF}_0$
times a number. The mathematical details can be found in Appendix~B.

Afterwards a more complete calculation of the form factor 
will be developed. At the moment only the absorptive diagrams 
have been included and only the imaginary part is under
control. Moreover, let us consider the simplest case of a single resonance
exchange. The factor  $\frac{1}{1-\mbox{tr}(\mM)}$ together with the initial  
$\vec{\mF}_0$  generates a complex denominator $M_{V_1}^2-q^2-\xi(q^2)$:  a 
non-controlled real part  plus a well defined imaginary term, given by 
\bel{eq.anchoV}
\mbox{Im} \ \xi(q^2) \ \ = \ \ \mbox{Im} \ \xi_\pi (q^2) 
+ \mbox{Im} \ \xi_K(q^2) \ .  
\ee
The bubble loop summation provides an imaginary contribution 
which gets separate contributions from the $\pi\pi$ and 
$K\overline{K}$ channels. The corresponding partial widths are 
provided by 
\be
\mbox{Im} \ \xi_P (q^2) \ = \ C_{P} \;\; 
\left(M_{V_1}^2\, -q^2+\frac{2\, G_{V_1}^2}{f^2}\, q^2 \right)
\frac{ q^2\sigma_P^3}{96\,\pi\, f^2}
\;\; \theta(q^2-4m_P^2) \,  ,
\ee
with $C_{\pi}=1$ and $C_{K}=\frac{1}{2}$.
When we substitute the coupling at LO in $1/N_C$, $G_{V_1}=f/\sqrt{2}$,  
these imaginary terms Im $\xi_P(q^2)$ agree with the partial widths 
$M_\rho \Gamma_\rho^{(P)}(q^2)$ obtained in
Refs.~\cite{GyP,anchura Jorge}
from a simplified single-channel analysis.
This energy dependence for the width was 
long ago considered by Gounaris and Sakurai  
from general arguments~\cite{Sakurai}. As well, they
had exactly the same logarithm in their work that the one 
which naturally appears
in our calculation of the absorptive contribution  
through the Feynman integral $B_{22}^{(P)}$.

The correlator of two vector currents  and the 
$I=J=1$ partial-wave scattering amplitude, can be computed in a similar way.  
For the correlator we begin with a current effective vertex, 
like for the form factor, and  connect it to a $N$-loop final-state 
interaction, which ends into another current effective vertex. 
For the scattering amplitude  we start from a scattering effective 
vertex and go on connecting loops and scattering vertices in the same way. 
A similar rescattering effect appears in the three quantities.
The resulting ($s$--channel) $I=J=1$  scattering amplitude takes the form:
\bel{eq.scatDyson}  
T \ = \ \Sigma \left(\,\sum_{N=0}^\infty\; \mM^N 
\right) \Sigma^{-1}\;
T_{\mbox{\tiny LO}}^s \; =\; 
\Sigma \ \left( 1 - \mM \right)^{-1}
\Sigma^{-1} \; T^s_{\mbox{\tiny LO}}\; =\; 
\Frac{1}{1-\mbox{tr} \{ \mM \} }\;\; T_{\mbox{\tiny LO}}^s\, .
\ee

The matrix structure $(1-\mM)^{-1}$ only depends on the scattering 
effective vertex and on the two-intermediate particle loop. 
As these are identical for the three quantities (VFF, correlator and
scattering), the 
final-state interaction dresses the bare resonance pole in a universal way,
providing the same complex pole for all processes. 

\section{Low-Energy Matching Conditions}

\tab
All the former calculations must reproduce the QCD low-energy
behaviour provided by the $\chi$PT framework. This allows to fix
the polynomial ambiguities at a given order in the chiral expansion.
We can identify the momentum expansion up to $\cO(E^4)$ of the resummed
vector form factor \eqn{eq.V0aV} with the standard $\cO(E^4)$ $\chi$PT
calculation in the usual $\overline{MS}-1$ scheme~\cite{MesonFF}. 
At leading order in $1/N_C$, we have 
the well-known relation \cite{the role,PI:02}
\bel{eq:L9}
\left.L_9\right|_{N_C\to\infty}\; =\; \sum_i \; 
{F_{V_i} G_{V_i}\over 2\, M_{V_i}^2}\, ,
\ee
with $F_{V_i}, \ G_{V_i}$ and $M_{V_i}$ 
the bare parameters of the R$\chi$T lagrangian.

Keeping $1/N_C$ corrections, the $\cO(E^4)$ matching
determines the regularized function
$B_{22}^{r,(P)}$, up to the considered chiral order, to be
\cite{GyP,anchura Jorge,PP:01}
\bel{eq.matchingV}
B_{22}^{r,(P)} \; =\; 
 \Frac{1}{192\pi^2}\left[
 \sigma_P^3\,\ln{\left(\frac{\sigma_P+1}{\sigma_P-1}\right)}
+ \ln{\left(\frac{m_P^2}{\mu^2}\right)} - \Frac{5}{3}  
+ \Frac{8\, m_P^2}{q^2} \right] - \Frac{2}{3}\,\delta L_9^r(\mu)\, ,
\ee
where \
$\delta L_9^r(\mu) \equiv  L_9^r(\mu)\, -\,
\left. L_9\right|_{N_C\to\infty}$.
The renormalization scale dependence of the
$\cO(E^4)$ $\chi$PT coupling $L_9^r(\mu)$ cancels out with the term
$\ln\left(m_P^2/\mu^2\right)$. 
The resulting vector form factor from~\eqn{eq.V0aV} takes then the form
\bel{eq.FFunares0} 
\ba{ccl}
\vec{\mF}  \ \ &=& \ \  
\Frac{ 1 + \sum_i \frac{F_{V_i} G_{V_i}}{f^2} \frac{q^2}{M_{V_i}^2-q^2} }{  
1 + \left( 1 + \sum_i \frac{2 G_{V_i}^2}{f^2} \frac{q^2}{M_{V_i}^2-q^2}  \right)  
\frac{2 q^2}{f^2} \left[ B_{22}^{r,(\pi)} + \frac{1}{2} B_{22}^{r,(K)}\right] } 
 \ \left(\ba{c} 1 \\ -\frac{1}{\sqrt{2}} \ea\right) \, .  
\ea
\ee

With the information obtained from the VFF we obtain as well the ($s$--channel)
$I=J=1$ partial-wave scattering amplitude, 
\bel{eq.scatunares}
T  \ \ = \  \ 
\Frac{ \Frac{q^2}{96 \pi f^2} 
\left( 1 + \sum_i \frac{2 G_{V_i}^2}{f^2} \frac{q^2}{M_{V_i}^2-q^2}  \right)}{  
1 + \left( 1 + \sum_i \frac{2 G_{V_i}^2}{f^2} \frac{q^2}{M_{V_i}^2-q^2}  \right)  
\frac{2 q^2}{f^2} \left[ B_{22}^{r,(\pi)} + \frac{1}{2} B_{22}^{r,(K)}\right] }
\left( \ba{cc} \sigma_\pi^2 &-\Frac{\sigma_\pi \sigma_K}{\sqrt{2}}\\[10pt] 
 -\Frac{\sigma_\pi \sigma_K}{\sqrt{2}} &  
\Frac{\sigma_K^2}{\strut 2}
  \ea \right)\, ,
\ee
with $B_{22}^{r,(P)}$ being the same than in the VFF due to the optical theorem. 

\subsection{Scale Running}

\tab
When the low-energy matching was performed, the unfixed $\delta L^r_9(\mu)$
parameter  was left. It appeared as an extra constant in $B_{22}^{r,(P)}$.  
This also pointed  
out an ambiguity in the election of the scale and in the 
renormalization scheme, usually  $\overline{MS}-1$ but not the unique one. 
For simplicity we will analyze this feature in the single resonance case and
with the leading values of the couplings, $F_{V_1}= \sqrt{2}  f = 2 G_{V_1}$. 
In this situation the VFF, for instance, becomes 
\bel{eq.FFunares}   
\ba{ccl}
\vec{\mF}  &=&   
\Frac{M_{V_1}^2}{M_{V_1}^2-q^2 + \frac{2M_{V_1}^2 q^2}{f^2}
\left(B_{22}^{r,(\pi)}
+\frac12 B_{22}^{r,(K)} \right) 
} \ \left(\ba{c} 1 \\ -\frac{1}{\sqrt{2}} \ea\right)
\\ \\  &=&
\Frac{M_{V_1}^2(\mu)}{M_{V_1}^2(\mu)-q^2 + \frac{2M_{V_1}^2(\mu) q^2}{f^2}
\left. 
\left(B_{22}^{r,(\pi)}
+\frac12 B_{22}^{r,(K)} \right)\right|_{\delta L^r_9(\mu)=0} 
\ + \ \cO(q^2/N_C^2) 
} \ \left(\ba{c} 1 \\ -\frac{1}{\sqrt{2}} \ea\right)\, ,
\ea
\ee
where  
\bel{eq.MVrun}
M_{V_1}^2(\mu)= M_{V_1}^2 
\left( 1- \frac{2\,\delta L^r_9(\mu)\, M_{V_1}^2}{f^2} \right) \, .
\ee
The second line in \eqn{eq.FFunares} is easily obtained
by multiplying the numerator and denominator with the factor 
$( 1- 2\,\delta L^r_9(\mu)\, M_{V_1}^2/f^2 )$. 
After introducing
this definition, $\delta L^r_9(\mu)$ shuffles from $B_{22}^{r,(P)}$ to  
the parameter $M_{V_1}^2(\mu)$.
Thus instead of two independent constants, $\delta L^r_9(\mu)$ and  
$M_{V_1}^2$, 
we only have the combination  $M_{V_1}^2(\mu)$ replacing everywhere 
the parameter $M_{V_1}^2$. The term $\delta L^r_9(\mu)$ disappears from 
$B_{22}^{r,(P)}$, hence leaving in the regularized Feynman integral 
an explicit dependence on $\mu$. 
Moreover,  the use of 
$M_{V_1}^2(\mu)$ in~\eqn{eq:L9} allows us to recover the whole value 
of the $\chi$PT running coupling 
\bel{eq.L9run}
L^r_9(\mu)\ \ = \ \ 
\left. L^r_9\right|_{N_C\to \infty} +  \delta L^r_9(\mu)
\ \ \simeq \ \ 
\Frac{f^2}{2\, M_{V_1}^2(\mu)} \ \ , 
\ee
up to the considered order. Therefore the parameter $M_{V_1}^2(\mu)$ 
captures the right dependence of $L_9^r(\mu)$ 
on the renormalization scale. In our phenomenological analysis, we will adopt
the usual reference value $\mu_0=770$ MeV.  
Later on we will perform numerical studies at different scales $\mu$ and will
examine the corresponding values of $L_9^r(\mu)$ derived 
through~\eqn{eq.L9run}.   
The prescription of eliminating $\delta L^r_9(\mu)$ from $B_{22}^{r,(P)}$ 
is assumed in the following. 

When studying the experimental data, we will observe that the couplings 
$F_{V_1}$ and $G_{V_1}$ 
are not exactly the ones provided by the large--$N_C$ limit, but they
have small deviations. These parameters suffer also slight variations
when more than one resonance is taken into account. In that case,
the scale dependence does not go in such a straightforward way to 
the parameter $M_{V_1}(\mu)$ as we
have seen in~\eqn{eq.MVrun}, although the relation is still obeyed within a 
given accuracy. The other parameters are going to suffer very tiny modifications
with the scale but, at the precision of our study, they remain
like constants.

\section{Phenomenology}

\tab
We are going to analyze the experimental data for the vector form factor, 
which is much cleaner than the one from $\pi\pi$ scattering. 
The vector form factor can be
experimentally tested in the photoproduction of pseudoscalars from $e^+e^-$
annihilation or in $\tau$ decay.  
Although there are many data from $e^+e^-$~\cite{Amendolia,Novo2000}, we
have decided not to consider them, as we have not taken into account the 
$\omega$--$\rho$ mixing. We have studied the $\tau\to\nu_\tau 2\pi$ data 
from ALEPH~\cite{ALEPH}, which provides a 
covariance matrix to account for experimental error correlations. 
Similar data from CLEO~\cite{CLEO} and OPAL~\cite{OPAL}
are also available.

\begin{figure}[t!]
\begin{center}
\includegraphics[width=10cm,clip]{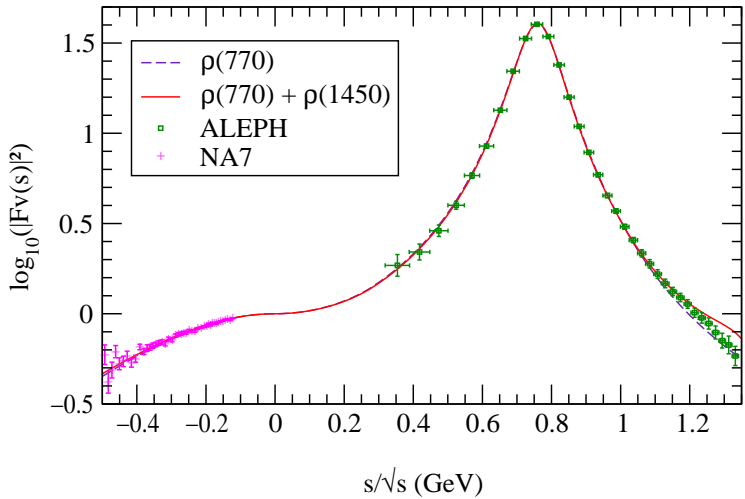}
\caption{\small{VFF fits to the $\tau\to\nu_\tau 2\pi$
ALEPH data~\cite{ALEPH}, with one
and two $\rho$ resonances. Also shown are the $e^+e^-\to 2\pi$
data points from NA7~\cite{Amendolia}.}}
\label{fig.dosrhoes}
\end{center}
\end{figure}

The range of validity up to which we will extend our fit is at most
$\sqrt{q^2}\leq 1.2$ GeV. Beyond this energy, multiparticle channels become
important.      
First we perform a fit to the modulus of the VFF (ALEPH data)  
with the $\rho(770)$ resonance only. This yields the 
parameter $M_{V_1}(\mu)$ and the couplings $F_{V_1}$ and $G_{V_1}$. 
We choose as matching scale $\mu_0=770$ MeV, take the pion decay constant 
$f=f_\pi=92.4$ MeV as an input, and fit the region 
$2\, m_\pi \leq \sqrt{s} \leq \Lambda_{\mbox{\tiny max}}=1.2$~GeV.
We obtain the values shown in Table~\ref{tab.FG}, 
with a $\chi^2/$dof$=24.8/25$. 
The corresponding VFF is shown in Fig.~\ref{fig.dosrhoes}.
In order  to estimate the systematic errors,  
we have varied the chiral parameter $f$ in the interval 
$f=92.4\pm 1.0$ MeV and the final point
of the fit $\Lambda_{\mbox{\tiny max}}$ between  
1.0 and 1.2 GeV. 
All these effects yield a more conservative result with a broader error. 
The first error in Table~\ref{tab.FG} is the one provided 
by MINUIT~\cite{MINUIT}, while the second is our estimated systematic 
uncertainty.

Besides the lagrangian parameter $M_{V_1}(\mu)$, we can determine 
the more usual ``physical''\, masses:
the Breit-Wigner mass $M_{_{BW}}$ and the pole mass 
$M_\rho^{\mbox{\tiny pole}}$.
The energy where the phase-shift $\phi_{\pi\pi}=\frac{\pi}{2}$ defines
the Breit-Wigner mass $M_{_{BW}}$ and the corresponding 
width is given by 
$1/\Gamma_{_{BW}} =  M_{_{BW}} 
\left. \frac{d\phi_{\pi\pi} }{ds}\right|_{s=M_{_{BW}}^2}$
\cite{Sakurai}. 
The complex pole of the observables in the second Riemann sheet,
$s^{\mbox{\tiny pole}}_\rho = (M^{\mbox{\tiny pole}}_\rho
- i\,\Gamma^{\mbox{\tiny pole}}_\rho /2)^2$, 
defines the alternative pole parameters.
%
%
%
%
%
%
%
\begin{table}
\begin{center}
\begin{tabular}{|c|c|c|}
\hline
\rule[-0.7em]{0em}{1.9em}
Chiral Coupling  &  $\rho(770)$  & $\rho(770)\ + \ \rho(1450)$ 
\\ \hline\rule[-1em]{0em}{2em}
$M_{V_1}(\mu_0)$    &   $845.4 \pm 1.1\, {}^{+0.8}_{-2.8}$ MeV 
&   $839.4 \pm 1.4\, {}^{+ 0.9}_{-2.3}$ MeV    
\\ \hline\rule{0em}{1.2em}
\rule[-1em]{0em}{2em}
$|F_{V_1}/f|$    &   $1.696 \pm 0.008\, {}^{+0.010}_{-0.028} $    
&    $1.669 \pm 0.008 \pm 0.017$  
\\ \hline\rule{0em}{1.2em}
\rule[-1em]{0em}{2em}
$|G_{V_1}/f|$  &  $0.695 \pm 0.004\, {}^{+0.011}_{-0.019} $   
&    $0.670 \pm 0.005 \, {}^{+0.012}_{-0.016}$ 
\\ \hline\rule{0em}{1.2em}
\rule[-1em]{0em}{2em}
$F_{V_1}G_{V_1}/f^2$  &   $1.178 \pm 0.010\, {}^{+0.009}_{-0.004}$   
&    $1.119\pm 0.012 \, {}^{+0.008}_{-0.018}$     
\\ \hline\rule{0em}{1.2em}
\rule[-1em]{0em}{2em}
$L_9^r(\mu_0) = \sum_i\frac{F_{V_i} G_{V_i}}{2 M_{V_i}^2(\mu_0)}$   
&  $( 7.04 \pm 0.05\, {}^{+0.19}_{-0.27} )  \cdot 10^{-3}$   
&   $(  6.79 \pm 0.09\, {}^{+0.19}_{-0.27} )  \cdot 10^{-3}$ 
\\ \hline\rule{0em}{1.2em}
\rule[-1em]{0em}{2em}
$M_{_{BW}}$ 	& $776.0 \pm 1.6\, {}^{+0.3}_{-0.7}$ MeV  &    
$773.9 \pm 2.0\, {}^{+0.3}_{-1.0}$ MeV
\\ \hline\rule{0em}{1.2em}
\rule[-1em]{0em}{2em}
$\Gamma_{_{BW}}$ 	& $156.2 \pm 1.6\, {}^{+0.3}_{-3.0} $ MeV  &   
$150.2 \pm 2.0\, {}^{+0.7}_{-1.6} $ MeV
\\ \hline\rule{0em}{1.2em}
\rule[-1em]{0em}{2em}
$M_\rho^{\mbox{\tiny pole}}$ 	
& $764.1 \pm 2.7\, {}^{+4.0}_{-2.5}$ MeV  &    
$770 \pm 3 \pm 3 $ MeV
\\ \hline\rule{0em}{1.2em}
\rule[-1em]{0em}{2em}
$\Gamma_\rho^{\mbox{\tiny pole}}$ 	
& $148.2 \pm 1.9 \, {}^{+1.7}_{-5.0} $ MeV  &   
$137.3 \pm 2.6 \pm 2.6 $ MeV
\\
\hline\end{tabular}
\caption{{\small  Determination of some R$\chi$T and $\chi$PT
couplings, at the
scale $\mu_0=770$ MeV, from the VFF fit. The parameters 
$F_{V_1}/f$ and $G_{V_1}/f$  have the same sign as 
$ F_{V_1} G_{V_1}/f^2>0$.}}
\label{tab.FG}
\vspace*{-1em}
\end{center}
\end{table}
%
%
In Table~\ref{tab.FG} we have written the resulting values
for these two different mass and width definitions.
In order to derive those numbers, we have taken into account
the correlations among the fitted parameters
$M_{V_1}(\mu_0)$, $F_{V_1}$ and $G_{V_1}$.
Owing to the off-shell $q^2$ behaviour of the denominator,
the pole mass turns out to be lower than the Breit-Wigner mass, in 
agreement with former works~\cite{pipicolangelo}. 
The opposite behaviour would have been obtained from a constant
Breit-Wigner width parameterization.  

\begin{figure}[t!]
\begin{center}
\includegraphics[width=10cm,clip]{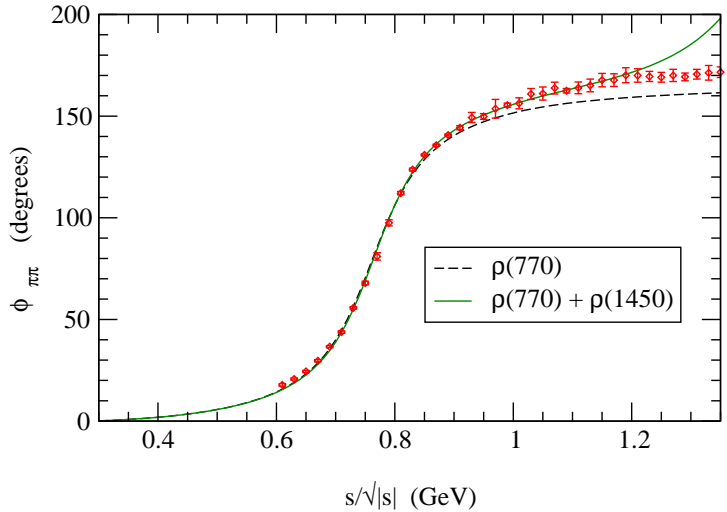}
\caption{\small{Phase-shift $\phi_{\pi\pi}$ of the $\pi\pi$ scattering
amplitude. The parameters employed for the one-resonance graph are the same 
than in Fig.~\ref{fig.dosrhoes}. The curve with two resonances takes
the values from the fit to the scattering amplitude, with the inputs 
$M_{V_2}(\mu_0)=1550$ MeV and $f=92.4$ MeV.
}}
\label{fig.fase}
\end{center}
\end{figure}

In Fig.~\ref{fig.fase} we plot the phase-shift
$\phi_{\pi\pi}$. In the low-energy region $\sqrt{s}\lsim 0.7$ GeV,
the experimental data appears to be slightly above the
predicted values. The same behaviour can be observed in previous theoretical
studies~\cite{GyP,PP:01,Palomar,IAM,N/D}.  
The experimental errors are probably underestimated in this region,
although higher-order chiral corrections could induce small
variations to our predictions.  
Other studies~\cite{pipicolangelo} seem to have a better control of the 
region closer to the $\pi\pi$ threshold and dominated by the $\chi$PT constraints. 
Beyond this region
the agreement of our one-resonance analysis with the scattering data 
is good up to  
$\sqrt{q^2}\leq 1$ GeV. Above this point the prediction for the scattering
amplitude  begins to fail.

In order to better study the region around $\sqrt{s}\sim 1$ GeV,
we include a second vector multiplet with the $\rho(1450)$.
The effect of the tail of the $\rho(1450)$ can modify
slightly the distribution in this region, where still the $\rho(770)$ dominates. 
Nonetheless, we cannot study energies much higher than $\sqrt{s}\sim 1.2$ GeV, 
since some not well-known strong inelasticities do arise
(the experimental phase-shift data does not seem to pass through
$3\pi/2$ at the $\rho(1450)$ mass~\cite{cuadernoCERN}).
Clearly, the two-pseudoscalar loops cannot incorporate all
the inelasticity needed to describe the $\rho(1450)$ region.
Other multiparticle intermediate states may be responsible for
this large effect.

We have fitted our theoretical
determination of the scattering amplitude with two resonances 
to the experimental phase-shift in the region
$0.7$ GeV$\leq \sqrt{s} \leq 1.2$ GeV.
The fit is not very sensitive to the $\rho(1450)$ mass, allowing
a wide range of values. Nevertheless, it requires that
$M_{V_2}(\mu_0)\gsim 1550$ MeV. Taking
$M_{V_2}(\mu_0)= 1550$ MeV, the fit to the phase-shift gives
$M_{V_1}(\mu_0)=841.8\pm 0.6$~MeV, \ 
$G_{V_1}/f=0.6631\pm 0.0027$ and $G_{V_2}/f=0.373\pm 0.028$, 
with $\chi^2/$dof$=18.8/22$.
The fitted value of $G_{V_2}/f$ increases for larger masses of the
$\rho(1450)$ resonance; the
central value grows to 0.57 for $M_{V_2}(\mu_0)= 2000$ MeV. 
The precision of $G_{V_1}/f$ is 
improved, as expected, because the phase-shift has a larger sensitivity
to this parameter.
The differences between the analyses of $\phi_{\pi\pi}$ with
one and two resonances are tiny for $\sqrt{s}\lsim 1$~GeV.
Beyond $\sqrt{s} \simeq 1.2$~GeV, the description breaks down 
because the pathological
$\frac{\pi}{2}(2n+1)$ behaviour of the phase-shift in the neighbourhood of
the $\rho(1450)$ still remains [see Fig.~\ref{fig.fase}].

We have performed next another fit to the VFF ALEPH data, with two vector 
multiplets and taking $\Lambda_{\mbox{\tiny max}}= 1.2$~GeV.
Since in this region the data have very small sensitivity to the 
$\rho(1450)$ mass and the coupling $G_{V_2}$, we introduce as an input
the value of $M_{V_2}(\mu_0)$ and the corresponding coupling $G_{V_2}$
obtained from the phase-shift fit.
The results of this VFF fit, given in Table~\ref{tab.FG}, 
have a $\chi^2/$dof$=14.7/24$.
The systematic errors have been estimated varying the pion decay constant
in the interval $f=92.4\pm 1.0$ MeV and
the value of $M_{V_2}(\mu_0)$ in the range\footnote{
Notice that the one-resonance results indicate that $M_{V}(\mu_0)$
is around 100~MeV larger than $M_{_{BW}}$ or $M^{\mbox{\tiny pole}}$.
The experimental situation of the $\rho(1450)$ is 
rather unclear and it might
be possible that it has an even higher mass or that a strong interference
of two vectors, $\rho(1450)$ and $\rho(1700)$,
is needed to properly describe the data~\cite{PDG}}
from $1550$ to
$2000$~MeV, which implies $G_{V_2}/f = 0.37\pm 0.03\, {}^{+0.2}_{-0.0}$.
In this analysis we have recovered as well the Breit-Wigner and
pole masses and widths for the $\rho(770)$ meson.
We have not tried to determine the $\rho(1450)$ pole, because it
would lie in a region which is not well described.
We also give in Table~\ref{tab.FG} the $\chi$PT coupling
$L_9^r(\mu_0)$ at the matching scale $\mu_0=770$ MeV.

The VFF fit is sensitive to the product of couplings
$F_{V_2} G_{V_2}/f^2$. One gets,
\be 
F_{V_2} G_{V_2}/f^2= 0.007\pm 0.024\, {}^{+0.000}_{-0.050} \, .
\ee
For the range of $G_{V_2}/f$ values quoted before, this implies
$F_{V_2}/f = 0.02 \pm 0.06\, {}^{+0.00}_{-0.08}$.

Modifications of the $\rho(1450)$ inputs produce sizable variations on the 
$\rho(770)$ couplings. Thus, a better knowledge of the $\rho(1450)$
is needed to get more accurate values of the $\rho(770)$ parameters
from a two-resonance fit. The results are consistent with the more precise 
determinations from the fit with only one resonance, which we take as our
best estimates.

\subsection{Running of $L^r_9(\mu)$}

\begin{figure}[t!]
\begin{center}
\includegraphics[width=10cm,clip]{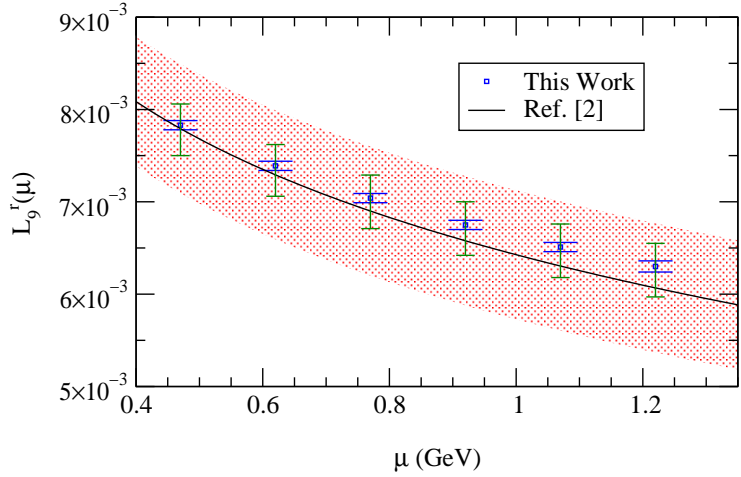}
\caption{\small{Comparison between the usually quoted
value of the chiral coupling $L_9^r(\mu)$~\cite{EFT} 
(shadowed band) and some determinations from  the fit at  
several matching scales: $L_9^r(\mu)=\sum_i \frac{F_{V_i}
G_{V_i}}{2M_{V_i}^2(\mu)}$. The smaller error intervals
are the statistical uncertainties given by MINUIT, the larger ones indicate
the total errors including systematic contributions.  
}}
\label{fig.L9}
\end{center}
\end{figure}
\tab
We have seen in Section 4, from a simplified theoretical analysis, 
that the parameter $M_{V_1}(\mu)$ depends on the $\chi$PT
renormalization scale adopted in the loop function $B_{22}^{r,(P)}$, 
in such a way
that the physically measurable VFF is scale independent as it should. 
The dependence of $M_{V_1}(\mu)$ with the scale was
given by the equation
\be
M_{V_1}^2(\mu_2) -  M_{V_1}^2(\mu_1) = \frac{M_{V_1}^2}{64\pi^2 f^2}
\ln{\left(\frac{\mu_2^2}{\mu_1^2}\right)} \, , 
\ee
as $L_9^r(\mu_2)-L_9^r(\mu_1)=\delta L_9^r(\mu_2)- \delta L_9^r(\mu_1)=
- \frac{1}{128 \pi^2}\ln{\left(\frac{\mu_2^2}{\mu_1^2}\right)}$. The 
theoretical running 
of $M_{V_1}(\mu)$ induces a scale dependence on the predicted  
value of $L_9^r(\mu)$ in Eq.~\eqn {eq.L9run}.
When the phenomenological fit is performed at different values 
of $\mu$, the parameter $M_{V_1}(\mu)$ increases with $\mu$. 
The other parameters of the fit remain essentially unaffected, i.e. they suffer
modifications much smaller than their errors.
Varying the scale $\mu$ in the range between 0.5 GeV and 1.2 GeV,  
the $\chi^2$ varies less than $2\%$.

The fitted $L_9^r(\mu)$ results are compared in Fig.~\ref{fig.L9} to 
the usually quoted values~\cite{EFT}.  
At the standard reference scale $\mu_0=770$ MeV, we obtain
\be 
L_9^r(\mu_0)= ( 7.04\pm 0.05\, {}^{+0.19}_{-0.27})\cdot 10^{-3} \, ,
\ee
which improves considerably previous determinations~\cite{EFT,BCT:98}.
The systematic errors would increase to $^{+0.19}_{-0.50}$ 
if we would have considered the fit with two resonances. 
The lack of knowledge about the second multiplet parameters
introduces an extra uncertainty of the same order than the one we 
have with only one resonance.

\subsection{Large--$N_C$ relations}

\begin{figure}[t!]
\begin{center}
\includegraphics[width=7cm,clip]{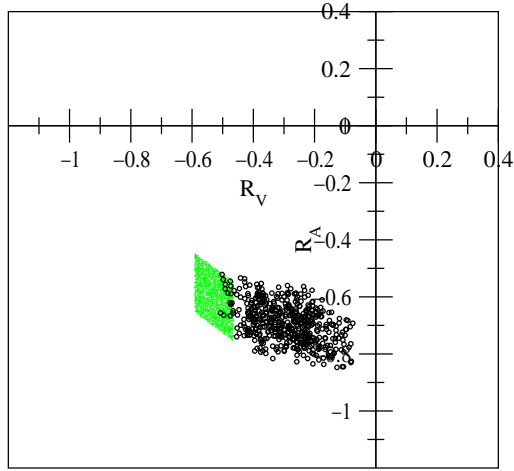}
\caption{\small{Allowed region in the $(R_V , R_A)$ plane.
The shaded zone on the left corresponds to the one-resonance study,
while the dotted zone results from the analysis with two resonances.
}}
\label{fig.NC}
\end{center}
\end{figure}
\tab
As we work at higher orders in $1/N_C$, our experimental
results have next-to-leading de\-via\-tions from the LO values 
provided by the two short-distance QCD relations~\eqn{eq:eps} 
and~\eqn{eq.delta}. We are going to test now  how well they are satisfied. 
Typically, there should be a deviation from zero of $\cO(1/N_C)$
in the VFF relation (with $N_C=3$ in physical QCD), as the
leading terms of the left-hand side of the equalities are 
$F_{V_1} G_{V_1}/f^2\sim 1$. The deviation in the axial form factor 
constraint should be of $\cO(1/N_C) \cdot 0.03$, because its leading 
terms are $2 F_{V_1}G_{V_1}/M_{V_1}^2\sim F^2_{V_1}/M_{V_1}^2\sim 0.03$.

In Fig.~\ref{fig.NC}, we have plotted the variables 
\be
\ba{ccl}
R_V &\equiv& 
\displaystyle\left\{ 1- \sum_i\; \Frac{F_{V_i} G_{V_i}}{f^2} \right\}  
\ \cdot \ N_C \ , \\ \\
R_A &\equiv&   \displaystyle\left\{  \sum_i\; 
\Frac{2 F_{V_i}G_{V_i}-F^2_{V_i}}{M_{V_i}^2} \right\}
 \ \cdot \ N_C \ (0.03)^{-1} \ , \\ 
\ea
\ee
which have been normalized with appropriate factors so that the 
expected deviations from zero are of $\cO(1)$. We have performed a scanning of
the range of values for the R$\chi$T couplings obtained from the VFF fits.  
We can see in the figure that 
the separation from the large--$N_C$ QCD relations is indeed of 
the expected order for both types of fits (with one or two resonances). 
Thus, the short-distance relations~\eqn{eq:eps} 
and~\eqn{eq.delta} are well satisfied, within the given accuracy.

\section{Uncertainties from Higher-Order Corrections}

\tab
There exist many more diagrammatic contributions which have not been
included in our results. We show in Appendix~C that, 
when the production of multiparticle states is neglected,
it is possible to define a generalized summation of Feynman
diagrams with two-body topologies. It makes use of a kernel function
$\mathcal{K}$, associated with the two-body scattering amplitude,
which incorporates those contributions not included in our effective
$s$--channel vertex of Fig.~\ref{fig.verteff}(b).
The resulting VFF can be formally written in a very compact form,
given in Eq.~\eqn{eq:finalVFF}.
Making a $1/N_C$ expansion of the kernel $\mathcal{K}$, one can easily check
that our $s$--channel result in Eq.~\eqn{eq.V0aV} corresponds to
the leading-order approximation.
The first correction originates from a single resonance exchange in
the $t$--channel, which induces a subleading contribution of 
$\cO(1/N_C)$ to the
kernel $\mathcal{K}$. The exchange of $n$ meson fields contributes to the
kernel at $\cO(1/N_C^n)$.

A general calculation of those higher-order corrections is a formidable
task. We know, however, that in the energy region we are studying
the tree-level scattering in the $t$--channel 
is much smaller than the one coming from the $s$--channel, what
seems to imply that they contribute as a small perturbation.
To estimate the size of those corrections, we have analyzed
the leading contribution from t--channel resonance exchange between
the final pions. According to the results in Appendix~C, it induces
a multiplicative correction into the VFF:
\bel{eq:VFF_t_app}
\vec{\mF} \;\approx\;  
\left[1- \mG^t_{1R}\right]^{-1} \left[ 1 
+\Sigma^{-1}T^s_{\mbox{\tiny LO}}\,\Sigma^{-1}(192\pi B_{22}^r) 
 \right]^{-1} 
\vec{\mF}_0\, ,
\ee
where $\mG^t_{1R}$ is the contribution from a single t--channel exchange.

The complete calculation of $\mG^t_{1R}$ is rather involved, since it
makes necessary to address the renormalization of R$\chi$T. This is a very
interesting issue, which we plan to analyze in a future publication
where a full analysis of the VFF at next-to-leading order
in $1/N_C$ will be attempted.
Here, we are only interested in its numerical impact on the results
presented in the previous sections. For simplicity, we will study
$\mG^t_{1R}$ in the SU(2) theory; i.e. we neglect the tiny contributions
from diagrams with kaons in the intermediate loop or in the final
state  
($\mG^{t \ (\pi, K)}_{1R}=\mG^{t \ (K,\pi)}_{1R}= \mG^{t \ (K,K)}_{1R}=0$).
Moreover, we will work in the chiral limit ($m_\pi=0$).

Although there are several Feynman diagrams contributing, we only need
to consider the dominant one where the current vertex 
$(k_1^\mu-k_2^\mu)$ produces a $\pi^-(k_1)\pi^0(k_2)$ pair, which is 
rescattered through a $t$--channel resonance.
This diagram generates the interesting non-analytic contributions,
plus a divergent local correction which should combine with the
local contributions from the other diagrams to provide a physical
finite result. Since we are interested in the region
$\sqrt{s} \ll 2 M_V$ (i.e. we work below the two-resonance cut), 
there are no additional sources of
non-analytic terms. The local ambiguity can be fixed to $\cO(E^4)$
by matching Eq.~\eqn{eq:VFF_t_app} with the known $\chi$PT result. This
requires $\mG^t_{1R}\sim \cO(E^4)$. The exchange of a vector resonance
can be easily computed in this way. One gets:
\be
\ba{ll}
\mG^{t \ (\pi,\pi)}_{1V}  = & 
\Frac{G_{V_1}^2\, M_{V_1}^2}{(4 \pi)^2 f^4}   \left\{
\left[ \mbox{Li}_2\left(1+\Frac{q^2}{M_{V_1}^2}\right)  
-\mbox{Li}_2\left(1\right) \right]\left( 2\, \Frac{M_{V_1}^4}{q^4} + 
5\,\Frac{M_{V_1}^2}{q^2}  + 2 \right)
\right. \\ \\ & \qquad\quad\;\left.
+\; \ln{\left(-\Frac{q^2}{M_{V_1}^2}\right)} 
\left(2\,\Frac{M_{V_1}^2}{q^2} + 4 
+ \Frac{\strut 1}{\strut 6}\,\Frac{q^2}{M_{V_1}^2}\right) 
  - 2\,\Frac{M_{V_1}^2}{q^2} - \Frac{\strut 9}{\strut 2} - 
 \Frac{\strut 35}{\strut 36}\,\Frac{q^2}{M_{V_1}^2} 
 \right\} \, .   
\ea
\ee

Since $\mG^{t \ (\pi,\pi)}_{1V} \sim q^4/M_{V}^2$,
we can neglect the exchange of higher-mass vector resonances.
However, we will also consider the $t$--channel
exchange of scalar resonances from the lightest multiplet,
with a mass $M_S\simeq 1$ GeV~\cite{the role, PDG} and couplings 
$c_d,c_m\simeq f/2$~\cite{kapi}.
It provides the contribution:
\bel{eq.ft1s}
\ba{ll}
\mG^{t \ (\pi,\pi)}_{1S}  = & 
\Frac{2\, c_d^2\, M_S^2 }{(4\pi)^2 f^4}\left\{ 
\left[ \mbox{Li}_2\left(1+\Frac{q^2}{M_{S}^2}\right)  
-\mbox{Li}_2\left(1\right) \right]
\left( 2\,\Frac{M_{S}^4}{q^4} + \Frac{M_{S}^2}{q^2}  \right)
\right. \\ \\ & \qquad\qquad\left. 
+ \ln{\left(-\Frac{q^2}{M_{S}^2}\right)} \left(2\,\Frac{M_{S}^2}{q^2} 
+  \Frac{\strut 1}{\strut 6}\,\Frac{ q^2}{M_S^2}\right) 
- 2\, \Frac{M_{S}^2}{q^2} - \Frac{\strut 1}{\strut 2} 
+ \Frac{\strut 1}{\strut 36}\,\Frac{q^2}{M_S^2} 
 \right\} \, .   
\ea
\ee
This result includes contributions from the singlet and the octet scalars.

At energies below and around the $\rho(770)$ peak, 
these $t$--channel diagrams give a correction smaller than 
$5\%$, which is within the uncertainties of the numerical analyses
performed in the previous section.
However, above $E\sim 1.2$ GeV the vector contribution
becomes larger than 10\% and these topologies cannot be neglected
any more. This kind of diagrams turn out to be very important at
high energies.

We have repeated our previous fits to the VFF ALEPH data, including 
the correction induced by
$\mG^{t \ (\pi,\pi)}_{1R} = \mG^{t \ (\pi,\pi)}_{1V} +
 \mG^{t \ (\pi,\pi)}_{1S}$.
The results of these fits are compatible with the ones obtained
before, showing that our former studies neglecting crossed channels
provide a good description within the given precision.

\section{Conclusions}

\tab
A quantum field theory description of strong interactions at
energies around the hadronization scale, $E\sim1$~GeV, requires
appropriate non-perturbative tools. While a fundamental
understanding of the confinement region of QCD is still lacking,
substantial phenomenological progress can be achieved through
effective field theories incorporating the relevant symmetries
and dynamical degrees of freedom.

Using an effective chiral lagrangian which 
includes pseudoscalars and explicit resonance fields,
we have investigated the VFF and related $I=J=1$ observables
in the interesting $E\sim1$~GeV energy range.
The heavy particles make the standard chiral counting in powers 
of momenta useless, because their masses
are of the same order than the chiral symmetry breaking scale.
Therefore, we have adopted instead the more convenient large--$N_C$ 
expansion, which provides a powerful tool to organize the calculation.

At the leading order in $1/N_C$, one gets an excellent description of the
VFF, far away from the resonance singularities.
A proper understanding of the zone close to the $\rho(770)$ pole,
requires the inclusion of next-to-leading contributions providing
the non-zero width of the unstable meson. The dressed propagator
can be calculated through a Dyson-Schwinger summation
of the dominant $s$--channel rescattering corrections,
constructed from effective Goldstone vertices containing both the
local $\chi$PT interaction and the resonance-exchange contributions
\cite{PI:02,GyP,anchura Jorge,PP:01}.

We have extended the Dyson-Schwinger summation of effective vertices
to handle problems with coupled channels in a sys\-te\-ma\-tic way,
through the recurrence matrix $\mM$. The inverse matrix
$\left(1-\mM\right)^{-1}$, generated by final-state interactions,
provides the right unitarity structure of
the observables~\cite{Palomar,IAM,N/D}.
Moreover, with an $SU(3)$--symmetric dynamics
(the vertices contain only derivatives and no quark masses),
$\left(1-\mM\right)^{-1}$ acts just like a pure number 
$\left[1-\mbox{tr}\{\mM\}\right]^{-1}$. Hence, there is no mixing among loops 
and the total decay width is simply given by a sum of separate
contributions from the different channels, which correspond
to the partial decay widths. 
An improved diagrammatic summation of more general two-body topologies 
has been given in Appendix~C. It includes the smaller
$t$--channel corrections, through the $1/N_C$ expansion of a non-trivial
interaction kernel $\mathcal{K}$ associated with the two-pseudoscalar
scattering amplitude.

The Feynman loops fully determine the non-analytic contributions, which are
dictated by unitarity and chiral symmetry. The local corrections, however,
are functions of the theoretically unknown couplings of the effective
lagrangian. They incorporate the short-distance dynamics and take care
of the regularization and renormalization prescriptions adopted in the
calculation. A significative reduction on the number of free parameters
is obtained, requiring the different amplitudes to satisfy the appropriate
QCD constraints at large momentum transfer \cite{the role,spin1fields}.
In fact, a very successful prediction of the most relevant $O(E^4)$ $\chi$PT 
couplings is obtained, under the reasonable assumption that the
lightest resonance multiplets give the dominant effects at low
energies \cite{PI:02}.
We have resolved the local ambiguities of the VFF,
imposing the QCD short-distance constraints and
performing a low-energy matching with the known $\cO(E^4)$ $\chi$PT result.

Working within the single-resonance approximation \cite{PI:02},
we have obtained a good fit to the ALEPH $\tau\to\nu_\tau 2\pi$
data~\cite{ALEPH}, in the range $2 m_\pi \leq \sqrt{q^2}\leq 1.2$~GeV.
At the chiral renormalization scale $\mu_0=770$ MeV, the fit
gives the values shown in Table~\ref{tab.FG} for the main $\rho$
parameters. The corresponding resonance pole 
$s^{\mbox{\tiny pole}}=(M^{\mbox{\tiny pole}}_\rho-i\,
\Gamma^{\mbox{\tiny pole}}_\rho/2)^2$ in the second Riemann sheet
is found to be at:
\be 
M_\rho^{\mbox{\tiny pole}} = 764.1\pm 2.7\, {}^{+4.0}_{-2.5}\;\mbox{MeV}\, ,  
\qquad\qquad 
\Gamma_\rho^{\mbox{\tiny pole}} = 148.2\pm 1.9\, {}^{+1.7}_{-5.0}\;\mbox{MeV}\, .
\ee 
We have achieved an improved determination of the $\chi$PT coupling 
\be 
L_9^r(\mu_0)= (7.04\pm 0.05\, {}^{+0.19}_{-0.27}) \cdot 10^{-3} \,  ,
\ee 
at $\mu_0=770$ MeV. Performing the phenomenological fit at several
scales $\mu$, ones obtains the proper running of $L_9^r(\mu)$ 
as prescribed by $\chi$PT.

To test the convergence of the $1/N_C$ expansion, we have analyzed the
deviations between the fitted parameters
and the corresponding theoretical large--$N_C$ predictions~\cite{spin1fields}. 
The differences are found to be of the expected $\cO(1/N_C)$ size, showing
that the limit $N_C\to\infty$ provides indeed an excellent description
of the local chiral couplings.

We have also investigated the corrections induced by the tail of
the $\rho(1450)$ vector resonance at the higher side of our energy range.
The effects are sizable, but the sensitivity is not good enough to
make a precise determination of its parameters or to disentangle the
existence of several higher-mass states. In order to do that, one
would need to study higher energies where other multiparticle final states,
beyond the two-body modes that we have analyzed, become relevant.
Moreover, a better calculation of $t$--channel contributions would be
needed, because they are no longer small above 1.2~GeV.

To summarize, we have performed a detailed analysis of the
$\rho(770)$ region, imposing all known theoretical constraints.
The main $\rho$ parameters and the $\chi$PT coupling $L_9^r(\mu)$
have been determined with rather good precision.
More work is needed to extend the results at higher energies.
It would also be very interesting to investigate in a similar way
the scalar sector, specially the pathological $I=J=0$ observables. 
We plan to address these issues in forthcoming works.

\section*{Acknowledgments}

We have benefit from many discussions with Jorge Portol\'es.
This work has been supported by MCYT, Spain (Grant FPA-2001-3031),
by EU funds for regional development
and by the EU TMR network EURODAPHNE (Contract ERBFMX-CT98-0169).

\newpage
\appendix
\newcounter{ap1}
\renewcommand{\thesection}{\Alph{ap1}}
\renewcommand{\theequation}{\Alph{ap1}.\arabic{equation}}
\setcounter{ap1}{1}
\setcounter{equation}{0}
\hspace*{-0.6cm}{\large \bf Appendix A: Feynman Integrals} 

\tab
The loop function $B_{22}^{(P)}$ used in the text is defined through
\bear
\Int \frac{dk^d}{i\, (2\pi)^d} \;
\frac{k^{\mu}k^{\nu}}{(k^2-m_P^2)\, \left[(q-k)^2-m_P^2\right]}
&\equiv& B_{22}^{(P)}\, q^2 g^{\mu\nu}\, +\, B_{21}^{(P)}\, q^{\mu}q^{\nu}\, ,  
\eear
with 
\be
B_{22}^{(P)} \; =\; \Frac{1}{192\pi^2} \left[ 
\left(1-\Frac{6 m_{P}^2}{q^2}\right)
\left[ \lambda_\infty
+\ln{\left(\frac{m_{P}^2}{\mu^2}\right)}\right]
 + \Frac{8m_{P}^2}{q^2}-\Frac{5}{3} 
+ \sigma_{P}^3 \ln{
\left(\frac{\sigma_{P}+1}{\sigma_{P}-1}\right) }  \right]  \ ,
\ee
where $\lambda_\infty\equiv \frac{2}{d-4}\,\mu^{d-4}
+\gamma_E - \ln{(4\pi)}-1$, $\gamma_E\simeq 0.5772$,
$\mu$ is the renormalization scale
and $\sigma_{P}\equiv\sqrt{1-4m_{P}^2/q^2}$
is the usual phase-space factor. 

The real part of this Feynman integral is divergent, but its imaginary parts 
is finite and takes the value
\be
\mbox{Im}\left\{ B_{22}^{(P)}\right\} \; =\; 
-\frac{\sigma_P^3}{192\pi} \;\theta(q^2-4 m_P^2) \, .
\ee

The dilogarithm function which arises in the crossed-channel calculations is
defined as
\be
\mbox{Li}_2(y) \; = \; - \ \Int_0^1\;
\Frac{dx}{x} \; \ln{(1-xy)}
\; = \; - \ \Int_0^y\;\Frac{dx}{x} \;\ln{(1-x)}\, . 
\ee
It has an imaginary part given by
\be
\mbox{Im}\left\{ \mbox{Li}_2(y+i\epsilon)\right\} 
\; = \; \pi \ln{(y)} \ \theta(y-1) \, .
\ee

\newcounter{ap2}
\renewcommand{\thesection}{\Alph{ap2}}
\renewcommand{\theequation}{\Alph{ap2}.\arabic{equation}}
\setcounter{ap2}{2}
\setcounter{equation}{0}
\hspace*{-0.6cm}{\large \bf Appendix B: Matrix Relations}

\tab
In the isospin limit, the matrix   
$\Sigma^{-1} T_{\mbox{\tiny LO}}^s\,\Sigma^{-1}$ 
is proportional to a dimension-one projector. 
Therefore, it obeys the properties of a general dimension-one 
projector $P$ and a general matrix~$B$:  
\be
P  \cdot  B \cdot  P  \;=\;\lambda\, P \, , 
\ee
with $\lambda=\ $tr$\left\{P \cdot B \right\}$.  
When the inverse matrix 
$\left(1-P \cdot B\right)^{-1}$ is multiplied  by the 
eigenvector $\vec{u}$ of $P$,  
or by the matrix $P$, we obtain
\bear 
\left(1-P \cdot B\right)^{-1} 
\  \vec{u} &=& \Frac{1}{1-\lambda} \ 
\vec{u} \, , \\
\left(1-P \cdot B \right)^{-1}
 \ P &=& \Frac{1}{1-\lambda} \ P \, . 
\eear
In the study carried on before in Section 3, the matrices $P$ and $B$ were
$\Sigma^{-1}T_{\mbox{\tiny LO}}^s\,\Sigma^{-1}$ and $(-192\pi B_{22})$, 
respectively. The matrix $P\cdot B$ was just  $\mM$ and   
the vector  $\vec{u}$  was  $\vec{\mF}_0$.   

\newcounter{ap3}
\renewcommand{\thesection}{\Alph{ap3}}
\renewcommand{\theequation}{\Alph{ap3}.\arabic{equation}}
\renewcommand{\thefigure}{\Alph{figure}}
\setcounter{ap3}{3}
\setcounter{equation}{0}
\setcounter{figure}{0}
\hspace*{-0.6cm}{\large \bf Appendix C: Summation
of General Two-Body Topologies}

\tab
The Dyson-Schwinger summation performed in Section 3 incorporates
the dominant s-channel contributions. Moreover, the adopted
matching procedure to the low-energy $\chi$PT results takes care of
tadpoles and local contributions, to the considered order in the
momentum expansion. There are, however, many more diagrammatic
topologies which have not been considered yet.
Neglecting the small corrections coming from multiparticle intermediate
states, it is possible to define a generalized summation of
Feynman diagrams with two-body topologies.

As we saw before, the effective vertex in Fig.~\ref{fig.verteff}(a) 
for the vector
current insertion producing a $P^- P^0$ pair of pseudoscalars
shows the momentum structure:
\bel{eq.generalVFF}
\vec{V}_0^{\mu} \; =\; (p_1-p_2)^{\nu} \, 
\left[ \  \vec{\mF}_0(s)\,  P_{T} {}_{\nu}^{\mu} \ 
+ \  \vec{\mF'}_0(s)\,  P_L {}_{\nu}^{\mu} \ \right] \sqrt{2}\, ,
\ee
with $P_T^{\mu\nu}=g^{\mu\nu}-\frac{q^\mu q^\nu}{q^2}$ and 
$P_L^{\mu\nu}=\frac{q^\mu q^\nu}{q^2}$ the usual transverse and longitudinal
Lorentz projectors.
In a similar way, the effective vertex in Fig.~\ref{fig.verteff}(b) describing
the $s$--channel scattering of two pseudoscalars, 
when projected on the P-wave ($I=J=1$), takes the form:
\bel{eq.generalVS}
\mT_0\; =\; (p_1-p_2)^{\beta} \
\left[ \Sigma^{-1} T^s_{\mbox{\tiny LO}}(s)\,\Sigma^{-1}\, P_T {}_{\beta}^{\alpha}  
\ +\   
\Sigma^{-1} {T'}^s_{\mbox{\tiny LO}}(s)\,\Sigma^{-1}\, P_L {}_{\beta}^{\alpha} 
 \right] \  \left(- \ \frac{48 \pi i}{q^2}\right) 
\  (k_1-k_2)_{\alpha} \ ,
\ee
with $p_1,\ p_2$ ($k_1, \ k_2$) the outgoing (incoming) momenta.
The matrix $T^s_{\mbox{\tiny LO}}$ is the corresponding 
$I=J=1$ partial-wave scattering amplitude.   

\begin{figure}[t!]
\begin{center}
\includegraphics[width=10cm]{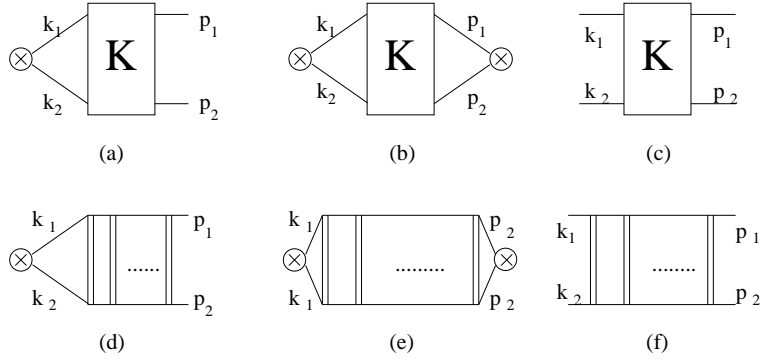}
\caption{\small{Basic pieces of the general summation of two-body topologies. 
The first row shows the general kernel $\mathcal{K}$ while the second one 
only includes the contributions from ladder diagrams.}}
\label{fig.krnl}
\end{center}
\end{figure}

Let us define a general kernel $\mathcal{K}^{(m,n)}(k_1,k_2,p_1,p_2)$
associated with the two-body scattering amplitude
from $(n)$--type pseudoscalars to $(m)$--type pseudoscalars.
This kernel, shown in Fig.~\ref{fig.krnl}(c), contains the identity
operator (no scattering) plus all interaction diagrams without
intermediate effective vertices \eqn{eq.generalVS}.

Now let us connect the effective vector current insertion
$\vec{V}_0^{\mu}$ to the kernel $\mathcal{K}^{(m,n)}$,
as shown in Fig.~\ref{fig.krnl}(a).
The outgoing pseudoscalars from the kernel 
are joined again into an effective scattering vertex \eqn{eq.generalVS}.
This generates the dressed structure:
\be
\ba{ll}
\vec{V}_1^{\mu} \; =& (p_1-p_2)^{\beta} 
\left[ \  \Sigma^{-1} T^s_{\mbox{\tiny L0}}(s)\, \Sigma^{-1}\,
P_T {}_{\beta}^{\alpha} \ 
+ \  
\Sigma^{-1} {T'}^s_{\mbox{\tiny LO}}(s)\, \Sigma^{-1}\,
P_L {}_{\beta}^{\alpha} \ \right] 
\ 
\left[\Pi(s)\, P_T {}_{\alpha}^{\nu} \ 
+ \  \Pi'(s)\, P_L {}_{\alpha}^{\nu}
\right]  
\\ \\
& \times \ 
\left[ \  \vec{\mF}_0(s)\,  P_T {}_{\nu}^{\mu} \ 
+ \  \vec{\mF'}_0(s)\,  P_L {}_{\nu}^{\mu} \ \right] \sqrt{2}   \, , 
\ea
\ee
with the matrices $\Pi(s)$ and $\Pi'(s)$ defined from the kernel integral
\bel{eq:PiDef}
\ba{l}
\Pi^{(m,n)}(s)\, P_T {}_{\alpha}^{\nu} \ +\   
\Pi'^{(m,n)}(s)\, P_L {}_{\alpha}^{\nu} \ \  
=  \ \ \left(-\Frac{48\pi i}{q^2}\right) \ \times \\ \\
 \; \Int \Frac{{dk'}^d}{(2\pi)^d} \Frac{dk^d}{(2\pi)^d}\,
({k'}_1-{k'}_2)_{\alpha}  
\Delta^{(m)}({k'}_1^2) \Delta^{(m)}({k'}_2^2)
\mathcal{K}^{(m,n)}({k'}_1,{k'}_2,k_1,k_2)  
\Delta^{(n)}(k_1^2)\Delta^{(n)}(k_2^2) 
 (k_1-k_2)^{\nu}  ,
\ea 
\ee
where $\Delta^{(m)}(k^2)$ is the propagator of a $(m)$--type
pseudoscalar.  
Performing the trivial products of Lorentz projectors, the 
vector current matrix element with
one intermediate kernel and ending into an effective scattering vertex 
takes the form:
\be
\vec{V}_1^{\mu}\; =\;   (p_1-p_2)^{\nu}
\left[\Sigma^{-1} T_{\mbox{\tiny LO}}^s(s)\, \Sigma^{-1}\,  \Pi(s) \,
\vec{\mF}_0(s) \,  
P_T {}_{\nu}^{\mu} \ 
+ \  \Sigma^{-1} {T'}_{\mbox{\tiny LO}}^s(s)\, \Sigma^{-1}  \ 
\Pi'(s)\,\vec{\mF'}_0(s)\, P_L {}_{\nu}^{\mu} \right] \sqrt{2}
\, \ .  
\ee
When the outgoing pseudoscalars are both 
on the mass shell the longitudinal term becomes zero.

We can easily iterate this algebraic procedure and consider a
series of $N$ intermediate kernels and effective scattering vertices, 
attached to the current insertion.
The first kernel is connected directly to $\vec{V}_0^{\mu}$;
then it comes an effective scattering vertex $\mT_0$, followed
by another kernel, and so on. The outgoing pseudoscalars are
attached to the final effective vertex.
The resulting contribution to the VFF is expressed as:
\be
\vec{V}_N^{\mu}\; =\; (p_1-p_2)^{\nu}
\left[ \ ( \Sigma^{-1} T_{\mbox{\tiny LO}}^s \Sigma^{-1}\Pi)^N 
\  \vec{\mF}_0 \  P_T {}_{\nu}^{\mu} \ 
+ \  ( \Sigma^{-1} {T'}_{\mbox{\tiny LO}}^s \Sigma^{-1} \Pi')^N
\  \vec{\mF'}_0\   P_L {}_{\nu}^{\mu} \ \right]  \sqrt{2}
\, .
\ee
Thus, the summation from $N=0$ to infinity becomes:
\be
\vec{V}^{\mu}  = (p_1-p_2)^{\nu}
\left[  ( 1 -\Sigma^{-1} T_{\mbox{\tiny LO}}^s \Sigma^{-1} \Pi)^{-1} \ 
\vec{\mF}_0 \  P_T {}_{\nu}^{\mu} \ 
+ \  (1-\Sigma^{-1} {T'}_{\mbox{\tiny LO}}^s \Sigma^{-1} \Pi')^{-1}
\  \vec{\mF'}_0\   P_L {}_{\nu}^{\mu} \ \right]   \sqrt{2} 
\, .   
\ee

This sums all diagrams ending in an effective scattering vertex. 
Finally, we add the diagrams where the last effective vertex is
connected to the outgoing pseudoscalars through the kernel. 
This extra contribution is given by
the form factor $\mG^\nu$  of the factorized element $(p_1-p_2)^{\nu}$,  
\be
\ba{rl}
\mG^\nu_{(m,n)} \; =&  (p_1-p_2)^{\beta} \, 
\left[\mG^{(m,n)}\, P_{T} {}_{\beta}^{\nu}+ 
\mG'^{(m,n)}\, P_{L} {}_{\beta}^{\nu}\right]  
 \; =   \\  \\ &  
\Int \Frac{dk^d}{(2\pi)^d} \; 
\mathcal{K}^{(m,n)}(k_1,k_2,p_1,p_2)\; 
\Delta^{(n)}(k_1^2)\,\Delta^{(n)}(k_2^2)\;
(k_1-k_2)^{\nu} 
\ ,
\ea
\ee
shown in Fig.~\ref{fig.krnl}(a), which  
we have separated into transverse and longitudinal parts.  
The summation of all types of diagrams gives then,
\be
\vec{V}^{\mu}  = (p_1-p_2)^{\nu}
\left[  \mG\, ( 1 -\Sigma^{-1} T_{\mbox{\tiny LO}}^s \Sigma^{-1} \Pi)^{-1} \ 
\vec{\mF}_0 \  P_T {}_{\nu}^{\mu} \ 
+ \  \mG'\, (1-\Sigma^{-1} {T'}_{\mbox{\tiny LO}}^s \Sigma^{-1} \Pi')^{-1}
\  \vec{\mF'}_0\   P_L {}_{\nu}^{\mu} \ \right]   \sqrt{2} \, .   
\ee

With the outgoing pseudoscalars being on-shell, the resulting VFF 
takes the compact form:
\bel{eq:finalVFF}
\vec{\mF} \; = \; 
\mG \cdot 
\left(1-
\Sigma^{-1}T_{\mbox{\tiny LO}}^s\,\Sigma^{-1} \Pi\right)^{-1} 
\cdot \,\vec{\mF}_0\, . 
\ee 

The simplest kernel is the trivial direct connection of the incoming and
outgoing pseudoscalars ($\mathcal{K}\,\dot=\, I$). In that case, the integral
\eqn{eq:PiDef} reduces to the usual two-propagator loop,
$\Pi = -192\pi\, B_{22}$, and $\mG = I$. One recovers then
the expression \eqn{eq.V0aV}, obtained through a Dyson-Schwinger summation
of s-channel scattering vertices.
Eq.~\eqn{eq:finalVFF} provides a systematic way of improving the result,
with the use of more complex kernels. The calculation could be
organized with the use of a $1/N_C$ expansion of the kernel $\mathcal{K}$;
the trivial identity operator corresponding to the lowest-order
approximation in this expansion. The first correction comes from
a single resonance-exchange in the $t$ channel, which induces a contribution
of $\cO(1/N_C)$ to the kernel. The exchange of $n$ meson fields would 
contribute at $\cO(1/N_C^n)$.

\subsection{Ladder Diagrams}

The calculation of higher-order diagrams with an arbitrary number
of resonances exchanged in the $t$ channel turns out to be a very 
complicated problem as each loop is connected to others.
However the optical theorem relates the form factor 
diagrams Fig.~\ref{fig.krnl}(d)  with  
the scattering amplitude through ladder diagrams, 
Fig.~\ref{fig.krnl}(f), in the familiar way~\cite{Palomar,IAM,N/D}:
\bear
\mbox{Im}\ T^t   &=&   
T^{t} \cdot \Sigma_\theta \cdot T^{t\, *} \, , 
\\
\mbox{Im}\ \mG^t   &=&   
\Sigma^{-1} T^{t}\,\Sigma^{-1} \cdot \Sigma_\theta^3 \cdot \mG^{t\, *}\, ,
\eear
implying
\bear
T^{t} &=& \left[T^{t\, -1}_{\mbox{\tiny LO}} + \cO(1) 
-i\,\Sigma_\theta\right]^{-1} \, , \\
\mG^t &=& \left[1 + \cO(1/N_C) 
- i \left( \Sigma^{-1}T^{t}_{\mbox{\tiny LO}}\,\Sigma^{-1} + 
\cO(1/N_C^2)\right) \,\Sigma_\theta^3\right]^{-1}\, , 
\eear
where the terms $\cO(1), \ \cO(1/N_C)$ and $\cO(1/N_C^2)$ correspond to NLO 
contributions in $1/N_C$, all of them real in
the physical region when multiparticle channels are neglected. 
The matrix $T^{t}_{\mbox{\tiny LO}}$ is the tree-level scattering
amplitude through a crossed resonance and the diagonal matrix $\Sigma_\theta$
is just the phase-space matrix but with each $\sigma_P$ multiplied by a
threshold factor $\theta(q^2-4 m_P^2)$.

The basic behaviour of these quantities is driven by the tree-level 
term, as the crossed scattering amplitude is tiny  at the energies we are 
considering.  It turns to be important 
at very high energy, where the $t$--channel becomes the dominant amplitude. 
Thus, the matching of $\mG^t$ to the lowest-order contribution
plus the diagrams with only one $t$--channel resonance exchange
is a suitable assumption:
\be
\mG^t \ \ \simeq \ \ \left[ 1 - \mG^t_{1R} \right]^{-1} \, ,
\ee
with Im $\mG^t_{1R}=\Sigma^{-1}T^t_{\mbox{\tiny LO}}\,\Sigma^{-1} \cdot
\Sigma_\theta^3$.

\end{document}